\documentstyle[12pt]{article}

\makeatletter
\def\lesssim{\mathrel{\mathpalette\vereq<}}
\def\vereq#1#2{\lower3pt\vbox{\baselineskip1.5pt \lineskip1.5pt
\ialign{$\m@th#1\hfill##\hfil$\crcr#2\crcr\sim\crcr}}}

\def\gtrsim{\mathrel{\mathpalette\vereq>}}
\makeatother

\newcommand{\beq}{\begin{equation}}
\newcommand{\eeq}{\end{equation}}

\newcommand{\remove}[1]{}

\begin{document}
\begin{titlepage}
\begin{center}
\today     \hfill    LBL-36338 \\

\vskip .5in

{\large \bf Possible Light U(1) Gauge Boson\\Coupled
to Baryon Number}\footnote{This work was supported by the
Director, Office of Energy Research, Office of High Energy and Nuclear
Physics, Division of High Energy Physics of the U.S. Department of Energy
under Contract DE-AC03-76SF00098.}


\vskip 0.5in

Christopher D. Carone  and Hitoshi Murayama\footnote{On leave of absence from
Department of Physics, Tohoku University, Sendai, 980 Japan.}

{\em Theoretical Physics Group\\
     Lawrence Berkeley Laboratory\\
     University of California\\
     Berkeley, California 94720}

\end{center}

\vskip .3in

\begin{abstract}
We discuss the phenomenology of a light U(1) gauge boson, $\gamma_B$,
that couples only to baryon number. We assume that the new
U(1) gauge symmetry is spontaneously broken and that the $\gamma_B$
mass is smaller than $m_Z$.  Nevertheless, we show that the model survives
the current experimental constraints.  In addition, we argue that evidence
for the existence of such a particle could be hidden in existing LEP and
Tevatron data.  We determine the allowed regions of
the $m_B$-$\alpha_B$ plane, where $m_B$ is the $\gamma_B$
mass, and where $4 \pi \alpha_B$ is the squared gauge coupling.  We point
out that in some parts of the allowed parameter space our model can
account for rapidity gap events in proton-antiproton scattering seen at
the Fermilab Tevatron.
\end{abstract}

\end{titlepage}
\renewcommand{\thepage}{\roman{page}}
\setcounter{page}{2}
\mbox{ }

\vskip 1in

\begin{center}
{\bf Disclaimer}
\end{center}

\vskip .2in

\begin{scriptsize}
\begin{quotation}
This document was prepared as an account of work sponsored by the United
States Government. While this document is believed to contain correct
 information, neither the United States Government nor any agency
thereof, nor The Regents of the University of California, nor any of their
employees, makes any warranty, express or implied, or assumes any legal
liability or responsibility for the accuracy, completeness, or usefulness
of any information, apparatus, product, or process disclosed, or represents
that its use would not infringe privately owned rights.  Reference herein
to any specific commercial products process, or service by its trade name,
trademark, manufacturer, or otherwise, does not necessarily constitute or
imply its endorsement, recommendation, or favoring by the United States
Government or any agency thereof, or The Regents of the University of
California.  The views and opinions of authors expressed herein do not
necessarily state or reflect those of the United States Government or any
agency thereof or The Regents of the University of California and shall
not be used for advertising or product endorsement purposes.
\end{quotation}
\end{scriptsize}

\vskip 2in

\begin{center}
\begin{small}
{\it Lawrence Berkeley Laboratory is an equal opportunity employer.}
\end{small}
\end{center}

\newpage
\renewcommand{\thepage}{\arabic{page}}
\setcounter{page}{1}

The standard model possesses a number of global U(1) symmetries that
are assumed to be accidental symmetries of the theory.  Baryon number,
and the three types of lepton number (associated with the electron, muon,
and tau) generate these U(1) symmetries.  It has been argued, however,
that global symmetries should be broken by quantum gravity effects
\cite{gravity}, with potentially disastrous consequences.  Baryon
number-violating operators generated at the Planck scale can lead to
an unacceptably large proton decay rate, especially in some
supersymmetric theories \cite{proton}.  This problem can be avoided
naturally if baryon number is taken instead to be a local symmetry.
Moreover, it is not even clear whether global phase rotations
are consistent with the basic premise of local field theory \cite{YM}.
For these reasons, and at the very least for aesthetics, it is natural
to wonder whether any of the global U(1) symmetries of the Standard Model
can be promoted to gauge symmetries in a phenomenologically acceptable way.

In this letter, we will consider the consequences of gauging the U(1)
symmetry generated by baryon number, U(1)$_B$.  We assume that the
symmetry is spontaneously broken and that the corresponding gauge
boson $\gamma_B$ develops a mass $m_B < m_Z$.  Of course, in the
minimal Standard Model we cannot gauge baryon number alone, because the
resulting field theory would suffer from gauge anomalies.
However, by adding a small number of new fermions
(that we can make heavier than $m_{{\rm top}}$ by an appropriate choice of
Yukawa couplings), we can gauge U(1)$_B$ in an anomaly-free way.  Then, the
main question of interest to us is whether the $\gamma_B$ boson could
have evaded all the available direct or indirect means of detection.  If
we call the squared gauge coupling $4\pi\alpha_B$, then we can determine
what regions of the $m_B$-$\alpha_B$ plane are excluded by the current
experimental constraints.  Considering the assumed lightness of the
$\gamma_B$ boson, our conclusions are somewhat surprising:  there are
relatively large regions of the $m_B$-$\alpha_B$ plane in which our
model is phenomenologically allowed \cite{ftnt}.  In addition, for some
of the allowed choices of the $\gamma_B$ coupling and mass, our model can
also account for the rapidity gap events observed at the Fermilab
Tevatron \cite{D0}.

We will first concern ourselves with the $\gamma_B$ phenomenology, and
then present an example of a simple, anomaly-free model at the end.
We will see that the $\gamma_B$ boson is elusive for some of the same
reasons that it is difficult to detect a light gluino \cite{gluino} or
stop \cite{stop}. Since the $\gamma_B$ boson couples only to quarks,
its most important effects can be expected in the same processes used in
measuring the QCD coupling $\alpha_s$.  Thus, we will determine the
allowed regions of the $m_B$-$\alpha_B$ plane by considering the
following observables: the $Z$ hadronic width, the $Z\rightarrow \mbox{3
jet}$ and $Z\rightarrow \mbox{4 jet}$ total cross sections, the di-jet
invariant mass distribution in $Z\rightarrow \mbox{4 jets}$, and the
hadronic decay width of the $\Upsilon(1S)$.  We will concentrate mostly
on the region where $m_B \gtrsim 10$ GeV, and the $\gamma_B$ boson
decays to $q\overline{q}$ with the width
$\Gamma_B = N_F \alpha_B m_B / 9$.  The more general case will be
considered in a longer publication \cite{carmur2}.

{\em The Z hadronic width.}  The $\gamma_B$ boson contributes to
the $Z$ hadronic width at order $\alpha_B$ through (1) direct production
$Z\rightarrow \overline{q}q\gamma_B$, and (2) the $Z\overline{q}q$
vertex correction.  Writing these two contributions as $F_{1}$ and $F_{2}$,
we find that the nonstandard contribution to the $Z$ hadronic width,
$\Delta \Gamma$, is positive and given by
\beq
\frac{\Delta\Gamma(Z\rightarrow \mbox{hadrons})}
{\Gamma(Z\rightarrow q\overline{q})} = \frac{\alpha_B}{18 \pi}
\left[F_{1}+F_{2}\right] ,
\eeq
where
\begin{eqnarray}
F_{1}&=&(1+\delta)^2 \left[3\ln\delta + (\ln\delta)^2\right]
+5(1-\delta^2)-2\delta\ln\delta
\nonumber \\
& &-2(1+\delta)^2
\left[\ln(1+\delta)\ln\delta+\mbox{Li}_2\left(\frac{1}{1+\delta}\right)-
\mbox{Li}_2\left(\frac{\delta}{1+\delta}\right)\right] , \\
F_{2}&=&-2\left\{\frac{7}{4}+\delta+(\delta+\frac{3}{2})\ln\delta \right.
\nonumber \\
& &\left. + (1+\delta)^2\left[\mbox{Li}_2\left(\frac{\delta}{1+\delta}\right)
+\frac{1}{2}\ln^2\left(\frac{\delta}{1+\delta}\right)
-\frac{\pi^2}{6}\right]\right\} .
\end{eqnarray}
Here Li$_2(x) = -\int_0^x \frac{dt}{t} \ln(1-t)$ is the Spence function,
and $\delta=m_B^2/m_Z^2$.  We compare this result to the uncertainty in
the experimentally measured $Z$ hadronic width corresponding to a two
standard deviation uncertainty in the extracted value of $\alpha_s(m_Z)
= 0.124\pm0.0086$ \cite{PDG}.  As shown in Fig. 1, this roughly excludes
the region of parameter space above $\alpha_B \approx 0.2$.

{\em Z $\rightarrow$ jets.} The $\gamma_B$ boson
contributes to $Z$ decay to four jets, via
$Z\rightarrow \overline{q}q\gamma_B$, $\gamma_B\rightarrow
\overline{q}{q}$.  In doing our parton-level jet analysis, we adopt the
JADE algorithm, in which we require jets $i$ and $j$ to be separated in
phase space by
$ y_{ij}\equiv 2E_iE_j(1-\cos\theta_{ij})/m_Z^2 > y_{{\rm cut}}$,
where $E_i$ and $E_j$ are the jet energies, and $\theta_{ij}$ is the
angle between the jets.  If any pair of jets has $y_{ij}<y_{{\rm cut}}$,
then these are combined into one jet, and the event instead contributes
to the three-jet cross section.  Since two of the jets originate from
the $\gamma_B$, the total four-jet
cross section as a function of $y_{{\rm cut}}$ will drop off
as $y_{{\rm cut}}$ is taken to be greater than $m_B^2/m_Z^2$.
The four-jet cross section is shown in Fig. 2 as a function of
$y_{{\rm cut}}$, normalized to the lowest order two-jet cross
section $\sigma_0$, for $\alpha_B=0.1$ and for a range of $m_B$
\cite{BASES}.  We compare our results to the experimental bounds on
the fraction of all four-jet events that are four-quark jet
events, 9.1\% (95\% C.L.) with $y_{{\rm cut}}=0.01$ \cite{OPAL}.
Comparing the $\gamma_B$ contribution to $\sigma_4/\sigma_0$ at
$y_{{\rm cut}}=0.01$ to the expected four-jet QCD background
$(\sigma_4/\sigma_0)_{\mbox{QCD}} \approx 0.2$ gives us the bound shown
at the top of Fig. 1.  For the most part, this excludes no new parameter
space beyond the region already excluded by our analysis of the $Z$
hadronic width.

The events that are not counted as four-jet events contribute to
the total three-jet cross section, in principle yielding some
enhancement over the expected rate.  However, given the large
three-jet QCD background, the three-jet analysis will not yield
a further constraint.

{\em Di-jet invariant mass peak in $Z\rightarrow 4\,\mbox{jets}$.} The
di-jet invariant mass $m_{jj}$ distribution in $Z$-decay has been studied
in searches for charged Higgs pairs, associated light and heavy Higgs
production in two-Higgs-doublet models, and excited quark pairs
that decay via $q^*\rightarrow q g$ \cite{associate}.  In these studies,
peaks in the $m_{jj}$ distribution from both particles were
required, so that the results are irrelevant to our problem.  In principle,
one can look for a peak in the $m_{jj}$ distribution without any other
requirements, but then one must contend with a huge QCD background. We
show the $m_{jj}$ distributions in Fig.~3 for various
values of $m_B$, together with the QCD background. We chose $y_{{\rm cut}} =
0.04$ to optimize the signal for $m_B = 20$~GeV.  It is clear that the
signal is overwhelmed by the background.  A distribution that is more
sensitive to the $\gamma_B$, especially for $m_B \lesssim 30$~GeV, is the
distribution of the smallest invariant mass  $m^2_{{\rm min}}$ among the
six possible combinations in four-jet events.  We show the
distribution of $y_{{\rm min}} = m^2_{{\rm min}} / m_Z^2$ in Fig.~4. The
background dominates the signal by more than a factor of 7, even on the
peak. Moreover, the peak will be further smeared by hadronization
effects and the resolution of the hadron calorimetry. Therefore no
practical constraint exists from the $m_{jj}$ distribution.
The search for a peak in the $m_{jj}$ distribution at hadron colliders
is probably hopeless, given the much larger backgrounds.

{\em $\Upsilon(1S)$ Decay.}    The decay of $\Upsilon(1S)$ is another
place to look for the effect of the $\gamma_B$ boson, through its
contribution to $R_\Upsilon =\Gamma (\Upsilon \rightarrow
\mbox{hadrons})/\Gamma(\Upsilon \rightarrow \mu^+ \mu^-)$.
The constraint that we obtain depends on whether $\gamma_B$ appears as a
real particle in the final state (when $m_B < m_\Upsilon$) or not
(when $m_B > m_\Upsilon$).  In the case where $m_B>m_\Upsilon$,
the most stringent constraint comes from the additional contribution to
the $\Gamma(\Upsilon \rightarrow \mbox{hadrons})$ from $s$-channel
exchange of the $\gamma_B$ boson. This additional
contribution is
\begin{equation}
\Delta R_\Upsilon = \frac{4}{3} \left[
	\frac{\alpha_B}{\alpha} \frac{m_\Upsilon^2}{m_B^2 - m_\Upsilon^2}
	+ \left( \frac{\alpha_B}{\alpha} \right)^2
	\left( \frac{m_\Upsilon^2}{m_B^2 - m_\Upsilon^2} \right)^2 \right],
\end{equation}
where $\alpha$ is the fine-structure constant.  This result includes
the interference with $s$-channel photon-exchange \cite{running}.  The
measured QCD coupling from $\Upsilon$ decay is $\alpha_s (m_Z) = 0.108
\pm 0.010$ \cite{PDG}, which implies $\Delta R_\Upsilon < 17.2$ at two
standard deviations. The resulting constraint on the free parameters
of our model is $m_B > m_\Upsilon \sqrt{1 + 43.8 \alpha_B}$ which is shown
in Fig 1. For the case where $m_B<m_\Upsilon$, the same argument gives us
the constraint $m_B < m_\Upsilon \sqrt{1 - 33.2 \alpha_B}$, also shown
in Fig. 1. For $m_B <m_\Upsilon$, another possible constraint comes from
the decay $\Upsilon \rightarrow \gamma_B g g $.  In the limit $m_B\approx 0$,
we obtain the limit $\alpha_B < 0.149$. Since this constraint is weaker
than those discussed above we have not shown it in Fig. 1.
The CP-even excited states $\chi_{b0}$, $\chi_{b1}$, $\chi_{b2}$ can decay
into a real $\gamma \gamma_B$ final state, so that a search for a
monochromatic photon may also exclude some portion of the parameter space.
This requires a more careful study, and will be discussed
elsewhere \cite{carmur2}.

If $m_B \gtrsim 20$~GeV, all analyses of $\alpha_s$ based on
deep inelastic scattering data, the lattice QCD calculations of
the quarkonia spectrum, and the $\tau$ hadronic decay width, will
remain unaffected by the existence of the $\gamma_B$. It is worth
pointing out that these measurements tend to give smaller values
of $\alpha_s$ compared to the value extracted from measurements made
at LEP, in particular, from the measurement of
$\Gamma(Z \rightarrow \mbox{hadrons})$. Since the $\gamma_B$ boson
provides an additional positive contribution to
$\Gamma(Z \rightarrow \mbox{hadrons})$, the data may be viewed as
suggesting its existence \cite{BD}. However, since the various measurements
of $\alpha_s(m_Z)$ seem to be converging, we feel that it is a more
conservative approach to restrict the parameter space of our model
based on the experimental data, rather than to predict specific
experimental anomalies.

Finally, we discuss signatures of the $\gamma_B$ boson that might be
discerned by further study of existent accelerator data.  In recent
analyses of four-jet events \cite{other}, the QCD group theory
factors $N_C$ and $T_F$ were fit using the $\theta_{BZ}$ and $\theta_{NR}$
distributions \cite{BZNR}.  The $\gamma_B$ contribution leads to an
enhancement in the number of $q\bar{q} q\bar{q}$ final states, similar
to the signature of an abelian gluon.  The fits allow $T_F$ to be
roughly twice as large as the QCD prediction, which would allow us to
exclude the region above $\alpha_B \simeq 0.1$ if $y_{{\rm cut}} \simeq
m_B^2/m_Z^2$ (see Fig.~3). However, the results in Ref.~\cite{other}
for $y_{{\rm cut}}= 0.01$--$0.03$ correspond to $m_B$ in the range
9--16 GeV, which is already excluded down to $\alpha_B \approx 0.04$
by the constraints from $\Upsilon$-decay. Thus, the data must be
re-analyzed for larger $y_{{\rm cut}}$ (up to $\approx 0.12$) before we
can put further constraints on the $m_B$-$\alpha_B$ plane.  The absence
of a next-to-leading order calculation of the QCD background, and the
lower statistics at higher $y_{{\rm cut}}$ will present the main
problems in this analysis.

Perhaps the most interesting potential signal of the $\gamma_B$ boson is
events with large rapidity gaps $\Delta \eta_c$ in hadronic collisions,
which are expected when scattering proceeds by color-singlet exchange.
At the Tevatron, rapidity gap events have been searched for at
$E_T > 30$~GeV and $\Delta \eta_c > 3$. In the large gap
limit ($\Delta \eta_c \gtrsim 4$), two-jet events are dominated by
$q\bar{q}$ scattering via gluon exchange because the center-of-mass energy
of the subprocess grows exponentially with
$\Delta \eta_c$, $\sqrt{\hat{s}} = 2 E_T \cosh \Delta
\eta_c/2$. The ratio of the events via $\gamma_B$ exchange to those by
gluon exchange is $(\alpha_B^2/ 18
\alpha_s^2) (1 + m_B^2/E_T^2)^{-2}$. Given an estimate of the survival
probability of the rapidity gap, $S \simeq 0.1$--$0.3$ \cite{survival}, the
contribution of $\gamma_B$ exchange to the rate of events with a
large rapidity gap is
\begin{equation}
f (\Delta \eta_c > 4) \sim (0.1\mbox{--}0.3)
	\frac{4\times 10^{-2}}{(1 + m_B^2/E_T^2)^2}
        \left( \frac{\alpha_B}{0.1} \right)^2.
\end{equation}
The rate is remarkably close to the experimental observations
\cite{D0} when $\alpha_B \simeq 0.1$ and $m_B \lesssim 30$~GeV.
While it has been suggested that the data could be explained by the
exchange of a QCD pomeron \cite{pomeron}, it is tempting to speculate
that $\gamma_B$ exchange might instead be the origin of the events
with large rapidity gaps \cite{HERA}.

Finally we present one simple extension of the Standard Model in
which U(1)$_B$ is gauged in an anomaly-free way.  To gauge the baryon
number current, we need to
introduce additional fermions to cancel
the U(1)$_B$ SU(2)$_L^2$,  U(1)$_B$ U(1)$_Y^2$,  U(1)$_B^2$ U(1)$_Y$
and U(1)$_B^3$ triangle anomalies. To do so, we introduce $N_Q$
families each consisting of an SU(2) doublet of left-handed
fermions $Q_L = (U_L, D_L)$ with zero hypercharge, and two SU(2) singlet
right-handed fermions, $U_R$ and $D_R$ with hypercharges $1/2$ and $-1/2$,
respectively. We assume that these new fermions acquire degenerate Dirac
masses from electroweak symmetry breaking (so that there will be no
contribution to the $T$ parameter). Assuming a common baryon number $B_Q$
for each of these fields, all anomalies cancel when $B_Q N_Q = -3$.
\cite{witten}. It is noteworthy that this particle content is exactly what
is found in a minimal one-doublet technicolor model \cite{technicolor},
with or without fundamental scalars \cite{carone}. The constraint from
the $S$-parameter \cite{peskin} $S_{new} = N_Q/(6\pi) < 0.46$
(95~\% C.L. \cite{PDG}) can be easily met when $B_Q \gtrsim 0.35$.

Since we have assumed that the $\gamma_B$ boson becomes massive through
spontaneous symmetry breakdown, there is also an associated Higgs boson.
However, since we do not know the Higgs boson's baryon number $B_H$, or
its quartic self-coupling $\lambda$, we cannot predict its mass $\sim
\lambda m_B/(\sqrt{4\pi \alpha_B}\, B_H)$.
Notice that if we let $B_H \rightarrow 0$, we can make the Higgs mass
arbitrarily large.  If we assume $B_H = 1/3$ and $\lambda \approx 1$,
then the mass of the Higgs boson will be around the 100~GeV scale. The
Higgs bosons decays into a real $\gamma_B \gamma_B$ pair, and thus, to
four jets.  It can be copiously produced by $\gamma_B$ fusion
in $q\bar{q}$ collisions or by the Bjorken-like process
$q \bar{q}\rightarrow \gamma_B^* \rightarrow \gamma_B H$, but the
final state is completely hadronic, and the signal is difficult to see.
It is important to note that the baryon number current is still conserved
even after the spontaneous breakdown of U(1)$_B$.  Therefore there is no
constraint from proton decay experiments.

{\it Conclusions.} We have shown that a new light U(1) gauge boson
$\gamma_B$ coupled to the baryon number is consistent with all existing
experimental constraints.  The allowed region of the model's parameter
space corresponds roughly to $m_B \gtrsim 20$~GeV
and $\alpha_B \lesssim 0.2$.  We have pointed out that the rapidity gap
events observed at Tevatron may be a manifestation of $\gamma_B$. We have
also shown that the gauge anomalies can be canceled easily by
introducing a small number of new fermions, with exactly the same
quantum numbers as in a minimal one-doublet technicolor model.

\begin{center}
{\bf Acknowledgments}
\end{center}
We are grateful to Mike Barnett, Lawrence Hall, Ian Hinchliffe, and
Axel Kwiatkowski for useful comments.
{\em This work was supported by the Director, Office of Energy Research,
Office of High Energy and Nuclear Physics, Division of High Energy
Physics of the U.S. Department of Energy under Contract DE-AC03-76SF00098.}



\newpage
\begin{center}
{\bf Figure Captions}
\end{center}

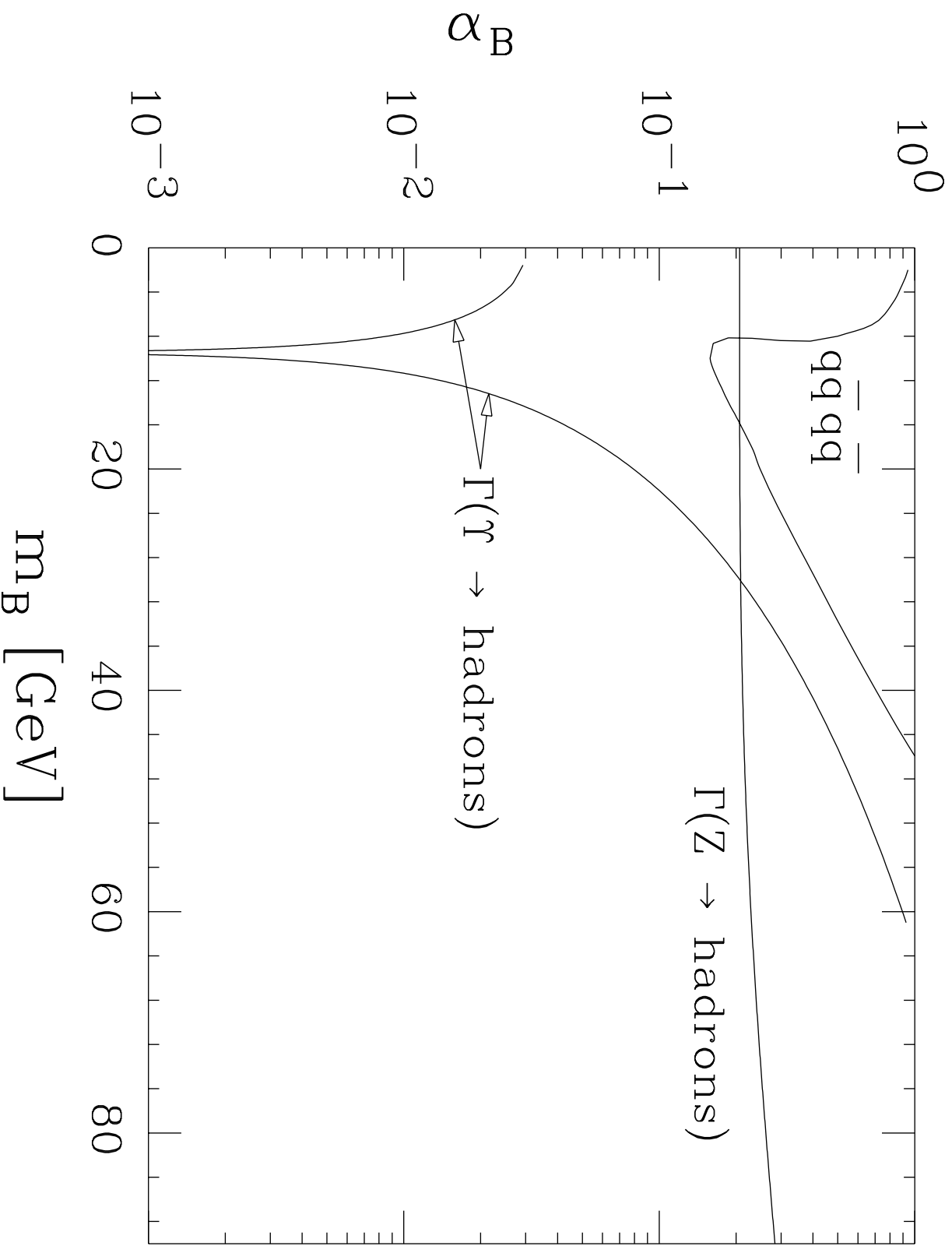
{\bf Fig. 1.}  Allowed regions of the $m_B$-$\alpha_B$ plane at 95~\%
C.L. or two standard deviations. The bounds shown come from (1) the $Z$
hadronic width, (2) the fraction of four-quark jet events in four-jet
events, (3) the hadronic decay width of the $\Upsilon(1S)$. The
parameter space above each of the lines shown is excluded, and the
weaker constraints discussed in the text are not shown.

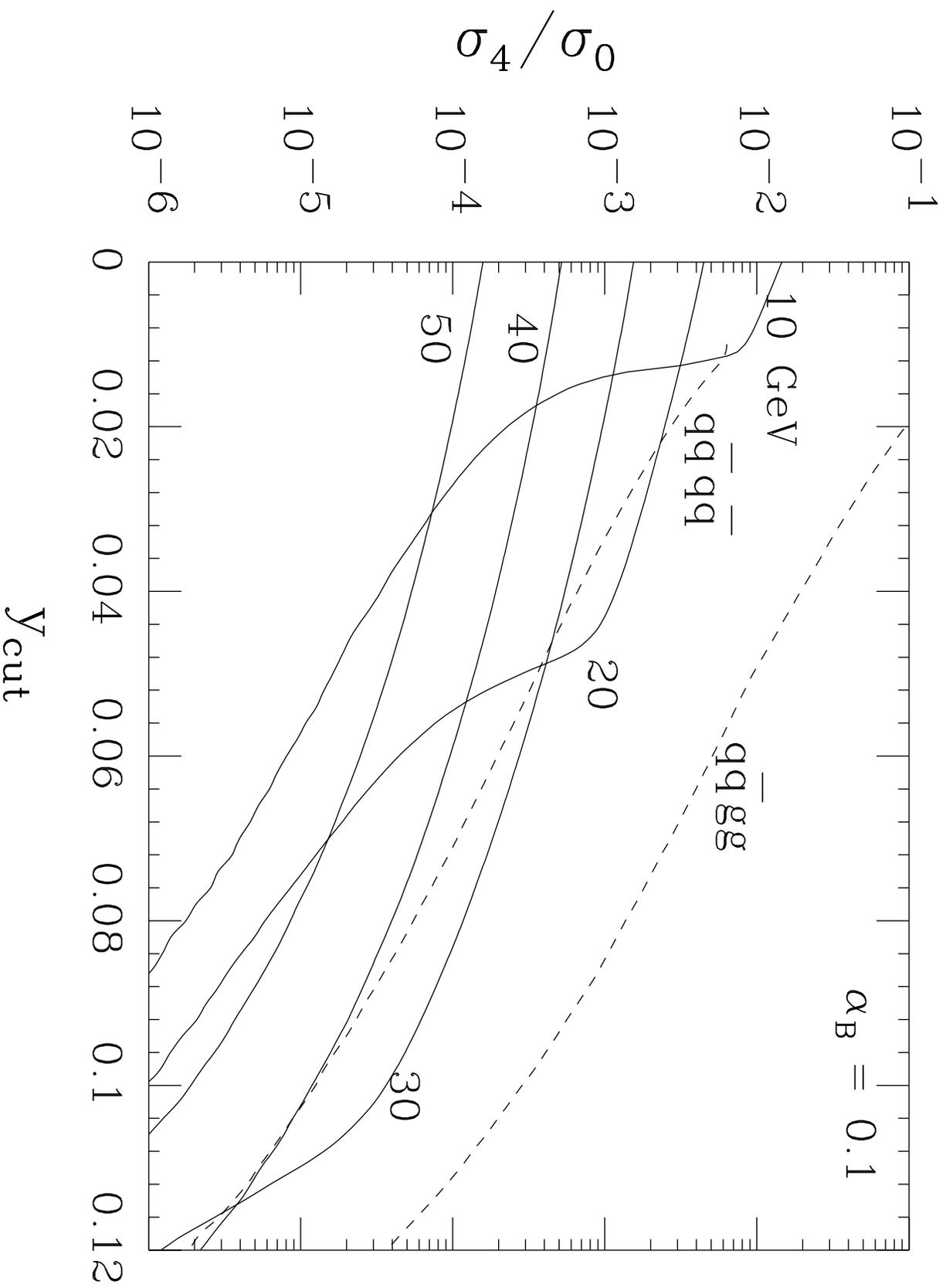
{\bf Fig. 2.} Four-jet cross section as a function of
$y_{{\rm cut}}$ for $\alpha_B=0.1$, normalized to the leading two-jet
cross section.

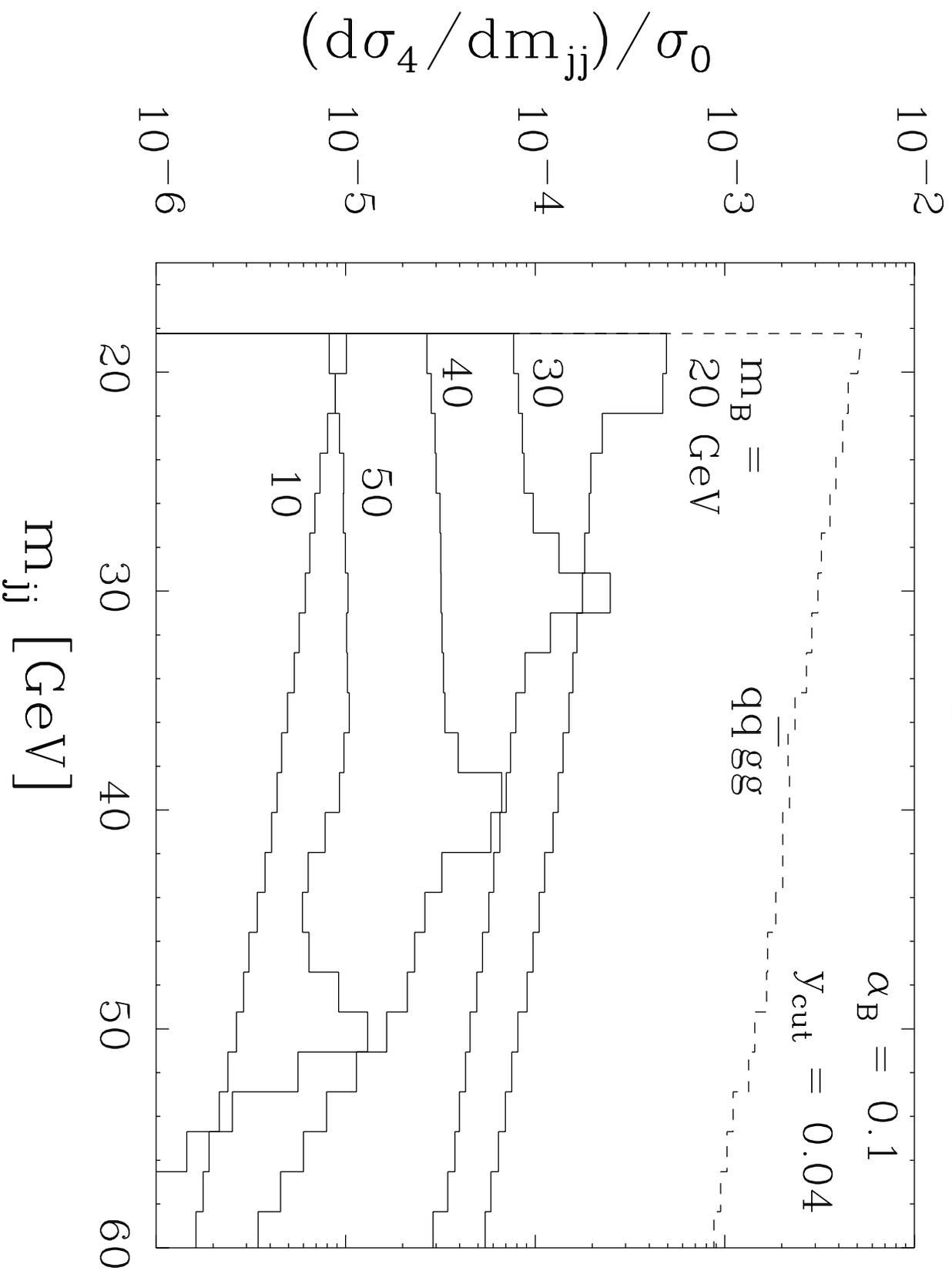
{\bf Fig. 3.} Di-jet invariant mass distribution in four-jet events,
for $\alpha_B=0.1$ and $y_{{\rm cut}}=0.04$, normalized to the leading
two-jet cross section.

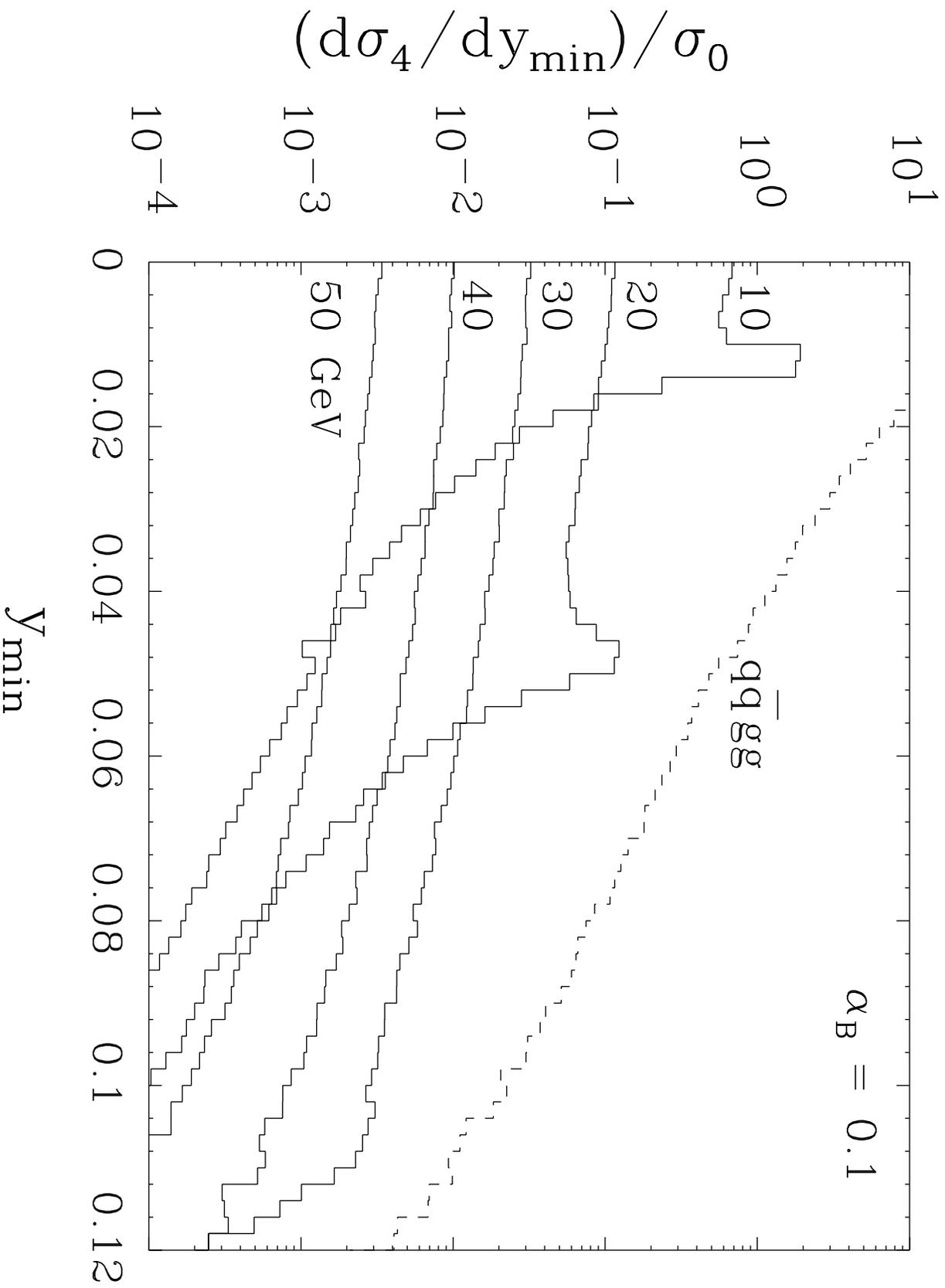
{\bf Fig. 4.} Four-jet differential cross section as a function
of $y_{{\rm min}}$ for $\alpha_B=0.1$, normalized to the
leading two-jet cross section.


\begin{thebibliography}{99}
\bibitem{running}
The running of $\alpha_B$ from $m_Z$ to $m_\Upsilon$ is small for
$\alpha_B \lesssim 0.2$, allowed by the $Z$ hadronic width.
\bibitem{HERA} For $m_B<m_\Upsilon$, there is also some
contribution to the rapidity gap events at HERA \cite{carmur2}.
\end{thebibliography}
\end{document}